\documentclass[prb,twocolumn,floatfix,amsmath,amssymb,showpacs,%
               superscriptaddress]{revtex4}

\usepackage{graphicx}
\usepackage{dcolumn}
\usepackage{bm}

\newcommand{\VC}[1]{{\bf #1}}

\newcommand{\tu}[1]{^{\textrm{#1}}}
\newcommand{\tl}[1]{_{\textrm{#1}}}
\newcommand{\MC}[1]{\mathcal{#1}}

\newcommand{\bohr}{\textrm{bohr}}

\newcommand{\fig}{figure}
\newcommand{\Fig}{Figure}
\newcommand{\tab}{table}

\newcommand{\bra}[1]{\langle#1|}
\newcommand{\ket}[1]{|#1\rangle}

\renewcommand{\Re}{\textrm{Re}}
\renewcommand{\Im}{\textrm{Im}}

\begin{document}
\title{Field-doping of C$_{60}$ crystals:
       Polarization and Stark splitting}
\author{Samuel Wehrli}
\email{swehrli@itp.phys.ethz.ch}
\affiliation{Theoretische Physik, ETH-H\"onggerberg, CH-8093 Z\"urich, 
             Switzerland}
\author{Erik Koch}
\affiliation{Max-Planck-Institut f\"ur Festk\"orperforschung,
             Heisenbergstra\ss e 1, 70569 Stuttgart, Germany}
\author{Manfred Sigrist}
\affiliation{Theoretische Physik, ETH-H\"onggerberg, CH-8093 Z\"urich, 
             Switzerland}
\date{\today}
\begin{abstract}
We investigate the possibility of doping C$_{60}$ crystals by applying a strong
electric field. For an accurate description of a C$_{60}$ field-effect device 
we introduce a multipole expansion of the field, the response of the C$_{60}$ 
molecules, and the Stark splitting of the molecular levels. The relevant 
response coefficients and splittings are calculated {\it ab initio} for several 
high symmetry orientations. Using a group theoretic analysis we extend these
results to arbitrary orientations of the C$_{60}$ molecules with respect to the
external field. 
We find that, surprisingly, for the highest occupied (HOMO) and the lowest
unoccupied molecular orbital (LUMO), respectively, the two leading multipole
components lift the degeneracy of the molecular level in the same way.
Moreover the relative signs of the splittings turn out to be such that the
splittings add up when the external field induces charge into the respective
level. That means that when charge carriers are put into a level, its 
electronic structure is strongly modified. Therefore, in general, 
in C$_{60}$ field-effect devices charge is not simply put into otherwise 
unchanged bands, so already because of this their physics should be quite 
different from that of the alkali-doped fullerenes.
\end{abstract}
\maketitle

\section{Motivation}

The proposal of doping C$_{60}$ crystals in a field-effect device (FET)
and the possibility of metallic conduction and even superconductivity in
such devices had raised wide-spread interest. While the
revelation of dishonest data handling in some cases\cite{beasley}
led to a severe damping of the initial enthusiasm, fundamental
aspects of field effect doping remains a timely and interesting
problem. For field-effect transistors made with self-assembled monolayers
this question was addressed in Ref.~\onlinecite{SAMFET} and for the reported
enhancement of the superconducting transition temperature in C$_{60}$ crystals
intercalated with haloform molecules in Refs.~\onlinecite{haloform}.
Attempts to observe the field-effect in graphite were reported in 
Ref.~\onlinecite{graphite}.
Here we address, from a theoretical point of view, the question how 
strongly C$_{60}$ can be doped in an electric field before its electronic
structure is substantially changed and how this structure changes
in even stronger fields. This is relevant for understanding the fundamental
features of field-effect devices based on C$_{60}$ and involves a number of
interesting physical problems.

It appears that doping  C$_{60}$ crystals in a field-effect device (FET) is
very hard to achieve in practice, one of the reasons being the exceptionally 
strong fields that  
are required. Very strong fields, however, not only induce charge carriers,
but also polarize the molecules, and, due to the Stark effect, in 
general, lift degeneracies. These effects are of particular importance, as 
C$_{60}$ is quite polarizable, and as its molecular levels are highly 
degenerate. The term ``field-doping'' naively implies that these effects are
small, such that the main effect of the external field is inducing charge
carriers into electronic levels which are essentially unaffected by the field.
It is clear that if the external field is strong enough, the electronic
structure of the crystal will be strongly modified by the field, and one thus
can no longer speak of doping. A fundamental question about field-effect
devices is therefore connected with the doping levels achievable, before
the electronic structure of the active material is substantially altered.
Moreover we discuss in detail how the electronic structure is changed, when 
the external field is beyond this ``doping limit''. In this regime the 
commonly used analogy of the field-doped C$_{60}$ with the alkali-doped 
fullerenes no longer holds. Nevertheless a field-effect device with high 
carrier concentration would be an interesting device in its own right. 
One might, e.g., speculate that for C$_{60}$ in a strong electric field the 
electron-phonon coupling is enhanced compared to unperturbed C$_{60}$, as 
for molecular orbitals of lower symmetry less couplings are forbidden by
symmetry.

For a first estimate of the fields involved in field-doping, consider a 
simple capacitor. Given a charge per area $\sigma$ on the plates, the 
electric field {\em between} the plates is $E=4\pi\sigma$. Assuming that in 
field-doped C$_{60}$ the induced charge resides in the top-most 
layer,\cite{art:wehrli1} inducing $n$ elementary charges per molecules 
requires an {\em external} field (originating from the gate electrode) of 
$E_\mathrm{ext}=2\pi n/A_\mathrm{mol}$, where 
$A_\mathrm{mol}$ is the area per molecule in the top-most layer of the
crystal. For a C$_{60}$ crystal with lattice constant $a\approx$~14 \AA\ 
typical areas per molecule are $A_{(111)}=\sqrt{3}a^2/4$ for the (111)-plane 
and $A_{(001)}=a^2/2$ for the (001)-plane. 
Even though these areas are quite large, the external fields necessary for
field doping are substantial, being of the order of 1 V/\AA\ per induced
elementary charge per molecule.
This is, however, not the field experienced by a molecule. As C$_{60}$ is
highly polarizable ($\alpha\approx 83$ \AA$^3$), in the solid, the external 
field at the site of a molecule is screened by the polarization of the 
neighboring molecules: A monolayer of dipoles $p$ centered on the lattice sites
$R_i$, generates a field $E_\mathrm{dip}= -p\sum_{i\ne0} |R_i-R_0|^{-3}$ at 
$R_0$, where the sum is over all sites in the monolayer, except $R_0$. 
For the (111) layer, this sum is about $31.2/a^3$, for the less dense (001) 
layer it is about $25.6/a^3$. As the dipole moments $p$ of the molecules are
induced by the screened field $E_\mathrm{scr}=E_\mathrm{ext}+E_\mathrm{dip}$
at the site of the molecule ($p=\alpha E_\mathrm{scr}$), we find, by 
self-consistent solution, that the external field is reduced by about
a factor of two.   

Inducing charge and polarizing the molecules is, however, not the only effect
of the external field. In addition it also leads to a splitting of the
molecular levels --- the Stark effect. As the molecular orbitals of C$_{60}$
have a definite parity, a homogeneous field splits the levels only in
second order. Thus for low fields the splitting is quite small, but
increases quickly for larger fields. We can expect that the
splitting of the molecular levels disrupts the electronic structure of
the crystal only when it becomes of the order of the band width, which is
about $1/2$ eV in C$_{60}$. 
Calculations indicate that the splittings of the $T_{1u}$- and $H_u$-levels 
in a {\em homogeneous} field are surprisingly small, being less than $1/2$ eV 
up to $E_\mathrm{scr}\approx 2$ V/\AA\ (cf.~\fig~\ref{splitting}).
Given the crudeness of the argument, it thus seems entirely
feasible that doping of a few elementary charges per molecule could be
achieved before the electronic structure of the C$_{60}$ is substantially
changed.

In the argument above we have described the C$_{60}$ molecules as polarizable
points. It is then natural to refine the model by considering also
higher multipoles. Such a multipole expansion is particularly suitable for 
C$_{60}$ as the molecules are nearly spherical. The approach is then
the following: First we calculate the response of a C$_{60}$ molecule to
external multipole fields. Then we self-consistently solve the electrostatic
problem for a lattice of molecules in a homogeneous external field. This 
provides us with the multipole expansion of the screened field acting on each 
molecule in the solid. Given that screened field, we can then determine the
splitting of the molecular levels as a function of the induced charge.

The organization of the paper reflects this approach: In section \ref{sec:dft} 
we describe the density functional calculations for determining the multipole 
response and the Stark splitting for a C$_{60}$ molecule in a multipole field 
for several symmetrical configuration. Using group theory, in section 
\ref{sec:irresp}, the irreducible parameters for the multipole response are 
determined. This allows the calculation of the polarization for arbitrary 
configurations. In section \ref{sec:split} we give an analogous treatment for 
the Stark splittings and explicitly show how the splitting changes as 
the molecule is rotated relative to the external field.
In section \ref{sec:mol_in_layer} we use these ingredients to self-consistently 
solve for the screened field seen by a molecule in a charged monolayer. 
The splitting of the molecular levels in this self-consistent multipole field
and the effect of this splitting on the density of states is presented in
section \ref{sec:scsplit}.
Our conclusions are presented in section \ref{sec:concl}.
The methods for calculating the response and splitting for an arbitrarily
oriented external field from the results of the density functional calculations
that were performed only for special orientations are described in the
appendices. Appendix \ref{sec:rsh} gives an example of how to calculate the 
response of a molecule using the irreducible parameters derived in section 
\ref{sec:irresp}. In appendix \ref{sec:coupling} we derive the coupling 
matrices needed for the calculating the level splitting when the molecule is 
rotated in the external field. Finally, appendix \ref{sec:shtranslation} 
gives the derivation of the matrix describing the field generated by a lattice
of identical multipoles at the origin, which is needed for finding the
self-consistent electrostatic field.

\section{Response of a C$_{60}$ molecule}\label{sec:dft}

To determine the response of a C$_{60}$ molecule to external multipole fields
we have performed all-electron density functional calculations using
Gaussian-orbitals.\cite{NRLMOL} The basis set comprises 5s4p3d for
carbon\cite{basis} and we use the Perdew-Burke-Ernzerhof functional.\cite{PBE}

\begin{figure}
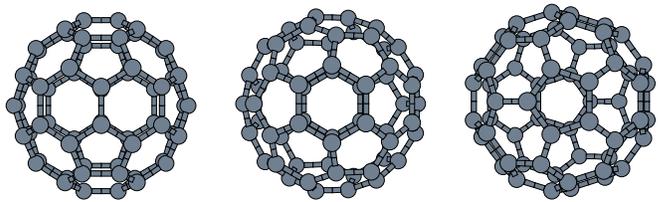

 \vspace{2ex}
 \resizebox{1in}{!}{\includegraphics{c60C2.epsi}}\hfill
 \resizebox{1in}{!}{\includegraphics{c60C3.epsi}}\hfill
 \resizebox{1in}{!}{\includegraphics{c60C5.epsi}}
 \caption[]{\label{orient} 
            Orientations of the C$_{60}$ molecule. The coordinate system
            is chosen such that the $x$-axis is pointing to the right, the
            $y$-axis upward, and the $z$-axis towards the reader. Thus,
            from left to right, the molecules are oriented with their
            2-, 3-, and 5-fold axis along $z$.
            The different orientations are obtained by rotating the
            molecule about the $y$-axis: by $\arctan(2-\tau)$ 
            to go from the first to the second, and by $\arctan(\tau)$
            to go from the first to the third orientation.
            $\tau=(\sqrt{5}+1)/2$ is the golden ratio.}
\end{figure}

\begin{figure}
 \resizebox{3.3in}{!}{\includegraphics{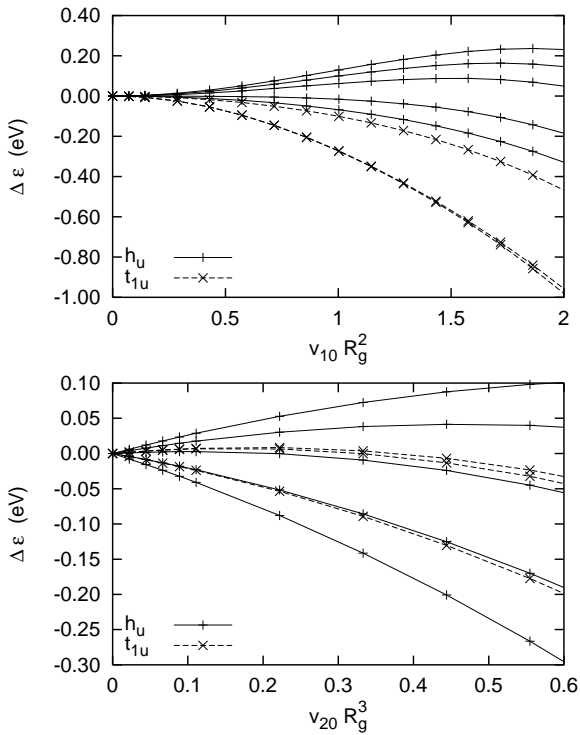}}
 \caption[]{\label{splitting}
           Splitting of the $H_u$ and $T_{1u}$ levels of a C$_{60}$-molecule in
           an external multipole field
           $V({\bf r})=V_{lm}R_{lm}^*({\bf r})$ 
           with $z$ along a 2-fold axis.
           Due to parity, for small external fields, the splitting for
           $(lm)=(10)$ is quadratic in the external field, while for
           $(lm)=(20)$ it is linear.
           The $V_{lm}$ are given in atomic units and $R_g=7$ \bohr.
           $(lm)=(10)$ corresponds to a homogeneous field with
           $E_z=-V_{10}$, thus $V_{10}\,R_g^2=1$ in atomic units corresponds
           to a homogeneous field of about 1~V/\AA.}
\end{figure}
In our calculations we apply an external multipole field and study the change
in the multipole moments of the charge density and the splitting of the
molecular levels as a function of the strength of the external field.
To take advantage of the molecule's symmetry we consider multipole fields
with the $z$-axis along the 2-, 3-, and 5-fold axis of the molecule
(cf.\ \fig\ \ref{orient}). As these axes are each contained in a mirror plane, 
which we chose to be the $x$-$z$-plane, we can treat the thus oriented molecule
as having symmetry group $C_{2v}$, $C_{3v}$, and $C_{5v}$ respectively. 
Applying external
multipole fields with $l>0$, this symmetry is maintained if the fields are 
proportional to the real part of the spherical harmonic $Y_{lm}$, where $m$ 
is an integer multiple of the order ($n=$2, 3, or 5) of the symmetry axis. 
Likewise, the response of the charge density will only have multipole 
components proportional to $\Re(Y_{lm})$, with $m$ an integer multiple of $n$. 
Calculations were done for such symmetry conserving multipole fields up to 
$l=6$. For the 3-fold axis oriented along $z$, we have, in addition, 
calculated the response to external fields proportional to $\Im(Y_{5,3})$, 
$\Im(Y_{6,3})$, and $\Im(Y_{6,6})$, i.e., with a symmetry lowered to $C_3$.
As we are interested in the linear response, we have considered small
multipole fields and made sure that the calculated response of the charge 
density is indeed proportional to the strength of the external field.
We find that the linear response of a C$_{60}$ molecule is very similar
to that of a metallic sphere of radius 4.4 \AA. This effective sphere radius
shows a slight increase with $l$. In addition there are weak off-diagonal 
terms.  To judge the accuracy of our calculation, we have checked how well the 
selection rules, that are not already imposed by the $C_{nv}$ symmetry, are
fulfilled for these off-diagonal terms. From the calculated 
response, we have determined the irreducible linear response coefficients, 
which will be given in the next section. 

In addition to polarizing the molecule, the external field also 
splits the degenerate molecular levels of the C$_{60}$ molecule. As the 
unperturbed molecular orbitals have a definite parity, for a multipole 
field with odd $l$ there will be no first-order splitting --- the quadratic 
Stark effect. On the other hand, a multipole field with even $l$ can
couple states of like parity, so in that case the splitting is linear. 
This is shown in \fig\ \ref{splitting}. In section \ref{sec:split} it will 
be demonstrated how the splitting of the HOMO- and LUMO-orbitals that were 
calculated for high-symmetry geometries can be extended by group theory 
to arbitrary 
orientations of the molecule relative to the external multipole field.

All calculations have been performed for the equilibrium geometry of the
unperturbed, neutral 
C$_{60}$ molecule. To estimate the effect of an external field
on the shape of the molecule, we have relaxed the structure in the presence
of homogeneous external fields of up to 1 V/\AA. We find only small
changes (up to about 0.005 \AA) in the lengths of the bonds (1.40 \AA\ for
the short and 1.45 \AA\ for the long bonds). Likewise, the polarizability
changes by less than 1.5\%.

Finally, we have calculated the total energy of the isolated C$_{60}$
ion as a function of its charge $q$ (spin unpolarized calculation with
relaxed geometries) and extracted the second order term $1/2\,U_0\,q^2$.
To compare with previous calculations,\cite{Mark} we find for the 
polarizability (multipole field with $l=1$) of the neutral
molecule $\alpha=$9.3 \bohr$^3$/atom, and $U_0=$3.2 eV.

\section{Irreducible response parameters}\label{sec:irresp}

The polarizability $\alpha$ of a molecule describes the linear dependence
of the induced dipole moment $\VC p=\alpha\,\VC E$ on the applied electric 
field $\VC E$.
For a multipole expansion, $\alpha$ becomes a matrix $\alpha_{l_1m_1\,l_2m_2}$ 
describing the response to all multipole fields.\cite{book:Stone}
To fix the notation (which follows Ref.~\onlinecite{book:Stone}) we briefly
review the definition of the multipole response matrix. 
 
The solutions of the Laplace equation $\VC \nabla^2\,V(\VC r)=0$ are given by
\begin{equation} \label{eq:V}
  V(\VC r)=V\tu e+V\tu i
          =\sum_{lm}V_{lm}\,R_{lm}^*(\VC r)+
                      Q_{lm}\,I_{lm}^*(\VC r),
\end{equation}
where the two terms denote the external potential ($V\tu e$), 
and the induced potential ($V\tu i$) due to a charge distribution
$\rho$ located around $\VC r=0$. Note that both, the Laplace equation as 
well as the expansion of $V\tu i(\VC r)$ into multipoles only holds for
$\VC r$ which lie outside the charge distribution. We have introduced the
regular and irregular spherical harmonics,\cite{book:Stone}
\begin{eqnarray} \label{eq:rs}
  R_{lm}(\VC r)&=&r^l\sqrt{\frac{4\pi}{2l+1}}\, Y_{lm}(\Omega), \\
  I_{lm}(\VC r)&=&\frac{1}{r^{l+1}}\sqrt{\frac{4\pi}{2l+1}}\, Y_{lm}(\Omega).
\end{eqnarray}
Special cases for the regular spherical harmonics are $R_{00}=1$ and $R_{10}=z$, 
hence, the external field, $V_{00}=V_0$ corresponds to a constant shift, 
and $V_{10}=-E_z$ is the $z$-component of the electric field.
For the irregular spherical harmonics we have $I_{00}=1/r$ and $I_{10}=z/r^3$,
thus for the induced potential, $Q_{00}=q$ gives the monopole charge 
while $Q_{10}=p_z$ is the dipole moment.
Generally, the coefficients $Q_{lm}$ are the multipole moments of the  
charge distribution $\rho$ 
\begin{eqnarray} \label{eq:vi}
  Q_{lm}&=&\int d^3r\,\rho(\VC r)\, R_{lm}(\VC r).
\end{eqnarray}
Decomposing the charge distribution $\rho=\rho_0+\Delta\rho$ into
the unperturbed charge density and the change in the charge density
due to the external potential, we obtain a decomposition of the 
multipole moments $Q_{lm}=Q^{0}_{lm}+\Delta Q_{lm}$.
Within linear response, the coefficients $\Delta Q_{lm}$ of the induced
multipole moments depend linearly on the coefficients $ V_{lm}$ of the
external potential, which defines the linear-response matrix 
$\alpha_{l_1m_1\,l_2m_2}$:
\begin{equation} \label{eq:lr}
  \Delta Q_{l_1m_1}=
  -\sum_{l_2m_2}\,\alpha_{l_1m_1\,l_2m_2} V_{l_2m_2},
\end{equation}
where the sign takes into account that the induced fields oppose the external 
fields. Then $\alpha_{1m_1\,1m_2}$ gives the dipolar response tensor,
while $\alpha_{00\,00}$ is the self-capacitance $U_0$.

The interaction energy of the molecule with the external potential is
$E\tl{ext}=\int d^3r\,\rho(\VC r)\,V\tu e(\VC r)$, which, using the previous
definitions, reduces to
\begin{equation} \label{eq:Eext}
  E\tl{ext}=\sum_{lm}V_{lm}\,Q^*_{lm} \,.
\end{equation}
Therefore, $V_{lm}$ and $Q_{lm}$ are pairs of conjugate
variables and the {\em total} energy of the molecule as a function of the
external field is given by
\begin{equation} \label{eq:Etot}
  E\tl{tot}=-\frac{1}{2}\sum_{l_1m_1\atop l_2m_2}\,
    \alpha_{l_1m_1\,l_2m_2}\, V_{l_1m_1}\,V^*_{l_2m_2}+
    \sum_{lm} V_{lm}\,(Q^{0}_{lm})^*.
\end{equation}
Since this is a quadratic form, we see that the matrix $\alpha$ is 
hermitian. We can make $\alpha$ real and symmetric by unitarily transforming
to a real basis (using $\sqrt{2}\,\Re(Y_{lm})$ and $\sqrt{2}\,\Im(Y_{lm})$
instead of $Y_{lm}$ and $Y_{l,-m}$ for $m\ne0$). 

The structure of the response matrix $\alpha$ depends on the symmetry of the 
molecule. For a metallic sphere of radius $R$ the response is isotropic, i.e.,
$\alpha$ is diagonal in the basis of the spherical harmonics:
$\alpha_{l_1m_1\,l_2m_2}=\delta_{l_1l_2}\,\delta_{m_1m_2}\,R^{2l_1+1}$.
Lowering the symmetry to icosahedral, $I_h$, introduces some anisotropy.
To understand the response matrix for C$_{60}$ we have to consider
how the irreducible representations (IR) of the rotation group $SO(3)$
split into IRs of the $I_h$ (see \tab~\ref{tab:ldecompI}). An external 
multipole field of angular momentum $l$ can only give rise to a response 
with angular momentum $l'$, if both IRs of the $SO(3)$ contain a common IR 
of the $I_h$. In particular, because
of parity, fields with even (odd) $l$ can only give rise to responses with
even (odd) $l'$. Furthermore, as the irreducible representations with
$l\le2$ are also irreducible with respect to the $I_h$, for $l\le2$ we have
$\alpha_{lm\,lm'}=\alpha_l\,\delta_{m m'}$. Thus, restricting the multipole
expansion to $l\le2$, the response of C$_{60}$ is isotropic, with
$\alpha_0\approx8.1\,\bohr$, $\alpha_1\approx556\,\bohr^3$, and
$\alpha_2\approx44100\,\bohr^5$.

For $l>2$ the space spanned by the spherical harmonics $Y_{lm}$ is no longer
irreducible with respect to the $I_h$. Thus we need to find linear combinations
of the spherical harmonics that span the irreducible representations of the
icosahedral group. We call them $Y_{lxk}$ where $l$ and $x$ denote the
IR of the $SO(3)$ and $I_h$, respectively, while the index $k$ labels the
functions within an irreducible representation of the $I_h$. If in the 
decomposition an IR $x$ should occur several times, we would have to 
introduce an additional multiplicity label. However, as can be seen from 
\tab~\ref{tab:ldecompI}, up to $l=6$ each IR appears at most once. 
We therefore suppress the multiplicity label here. Explicit expressions 
for the basis functions $Y_{lxk}$, can, e.g., be found in Ref.\ 
\onlinecite{book:Butler}, chapter 16, or Ref.\ \onlinecite{dresselhaus}, 
\tab~4.2.

In the new basis, the matrix $\alpha$ is built of blocks of diagonal matrices 
\begin{equation} \label{eq:alphastr}
  \alpha_{l_1x_1k_1\,l_2x_2k_2}=\alpha_{l_1l_2}(x_1)\,\delta_{x_1x_2}\,
           \delta_{k_1k_2}\,,
\end{equation}
where $\alpha_{l_1l_2}(x_1)$ constitute the minimal set of parameters, and, 
$\alpha$ being real symmetric, $\alpha_{l_1l_2}(x_1)=\alpha_{l_2l_1}(x_1)$.
The matrix elements of $\alpha$ were calculated up to $l=6$ using the
results of the density functional calculations described in
section~\ref{sec:dft}. The $\alpha_{l_1l_2}(x_1)$ are listed in
\tab~\ref{tab:lr}.   
From this minimal set of independent parameters we can determine the
response for arbitrary orientations of the C$_{60}$ molecule relative to 
the external multipole field. An example of how to do this is given in 
appendix~\ref{sec:rsh}.
The practical advantage of this procedure is clear: We only need to perform
density functional calculations for a number of highly symmetric
configurations, 
for which the numerical effort is much reduced. Using group theory the response
for arbitrary configurations can then be determined from these special cases.

Note that the group theoretical approach presented in this section is
particularly elegant in the case of a neutral molecule which has icosahedral
symmetry. Upon charging, orbitals become partially filled and the 
symmetry is reduced which leads to a higher number of irreducible response
coefficients. Furthermore, the symmetry of a charged molecule depends on how
the additional charge arranges in the degenerate orbitals.  This is a subtle
question in the case of an isolated C$_{60}$ molecule and involves 
Jahn-Teller effects and Coulomb interaction in
competition~\cite{art:Auerbach}. In the present 
work we restrict the analysis to neutral molecules. 
\begin{table}
\begin{tabular}{c|c}
  $l$ & $I_h$  \\
  \hline
  $0$ & $A_g$  \\
  $1$ & $T_{1u}$  \\
  $2$ & $H_g$   \\
  $3$ & $G_u\oplus T_{2u}$   \\
  $4$ & $G_g\oplus H_g$  \\
  $5$ & $T_{1u}\oplus T_{2u}\oplus H_u$ \\
  $6$ & $A_g\oplus T_{1g}\oplus G_g\oplus H_g$ \\
\end{tabular} \\
\caption{\label{tab:ldecompI}
   Decomposition of the irreducible representations (IR) of the rotation
   group $SO(3)$ into the IR of the icosahedral group $I_h\subset SO(3)$.}
\end{table}
\begin{table}
\begin{tabular}{c|@{\;}D{.}{.}{7}}
  &\multicolumn{1}{c}{${\alpha_{l_1l_2}(x)\over R_0^{l_1+l_2+1}}$}\\
  \hline
  $\alpha_{00}(A_g)$    &  1.019    \\
  $\alpha_{11}(T_{1u})$ &  0.990(0)    \\
  $\alpha_{22}(H_g)$    &  1.154(0)    \\
  $\alpha_{3 3}(G_{u})$  & 1.268(1) \\
  $\alpha_{3 3}(T_{2u})$ & 1.376(3) \\
  $\alpha_{4 4}(H_g)$    & 1.542(6) \\
  $\alpha_{4 4}(G_g)$    & 1.477(9) \\
  $\alpha_{2 4}(H_g)$    & 0.074(6) 
\end{tabular} 
\hspace{5ex}
\begin{tabular}{c|@{\;}D{.}{.}{7}}
  &\multicolumn{1}{c}{${\alpha_{l_1l_2}(x)\over R_0^{l_1+l_2+1}}$}\\
  \hline
  $\alpha_{5 5}(T_{1u})$ & 1.707(9)  \\
  $\alpha_{5 5}(T_{2u})$ & 1.430(4)  \\
  $\alpha_{5 5}(H_u)$    & 2.031(13) \\
  $\alpha_{1 5}(T_{1u})$ &-0.077(5)  \\
  $\alpha_{3 5}(T_{2u})$ & 0.039(6)  \\
  $\alpha_{6 6}(A_g)$    & 1.598(5)  \\
  $\alpha_{6 6}(H_g)$    & 1.209(21) \\
  $\alpha_{6 6}(G_g)$    & 1.964(2)  \\
  $\alpha_{6 6}(T_{1g})$ & 2.503  \\
  $\alpha_{2 6}(H_g)$    &-0.023(13) \\
  $\alpha_{4 6}(H_g)$    & 0.195(12) \\
  $\alpha_{4 6}(G_g)$    &-0.337(13)  
\end{tabular} 
\caption{\label{tab:lr}
  Linear response coefficients of a neutral C$_{60}$ molecule derived from
  the results of our density functional calculations. The matrix elements
  $\alpha_{l_1 l_2}(x)$ are normalized with $R_0=8.25\,\bohr$. 
  By comparing matrix elements determined from the response to potentials
  applied along the 2-, 3-, and 5-fold axis, we have determined, wherever
  possible, the uncertainties in the values of the matrix elements.
  The value of $\alpha_{0 0}(A_g)$ is given by the quadratic term of the
  change of the ground state energy upon charging of the molecule.  } 
\end{table}
%

\section{Level splitting}\label{sec:split}

\begin{figure}
\begin{center}
\begin{minipage}[b]{0.4\textwidth}
  \begin{center}
   \includegraphics[width=\textwidth]{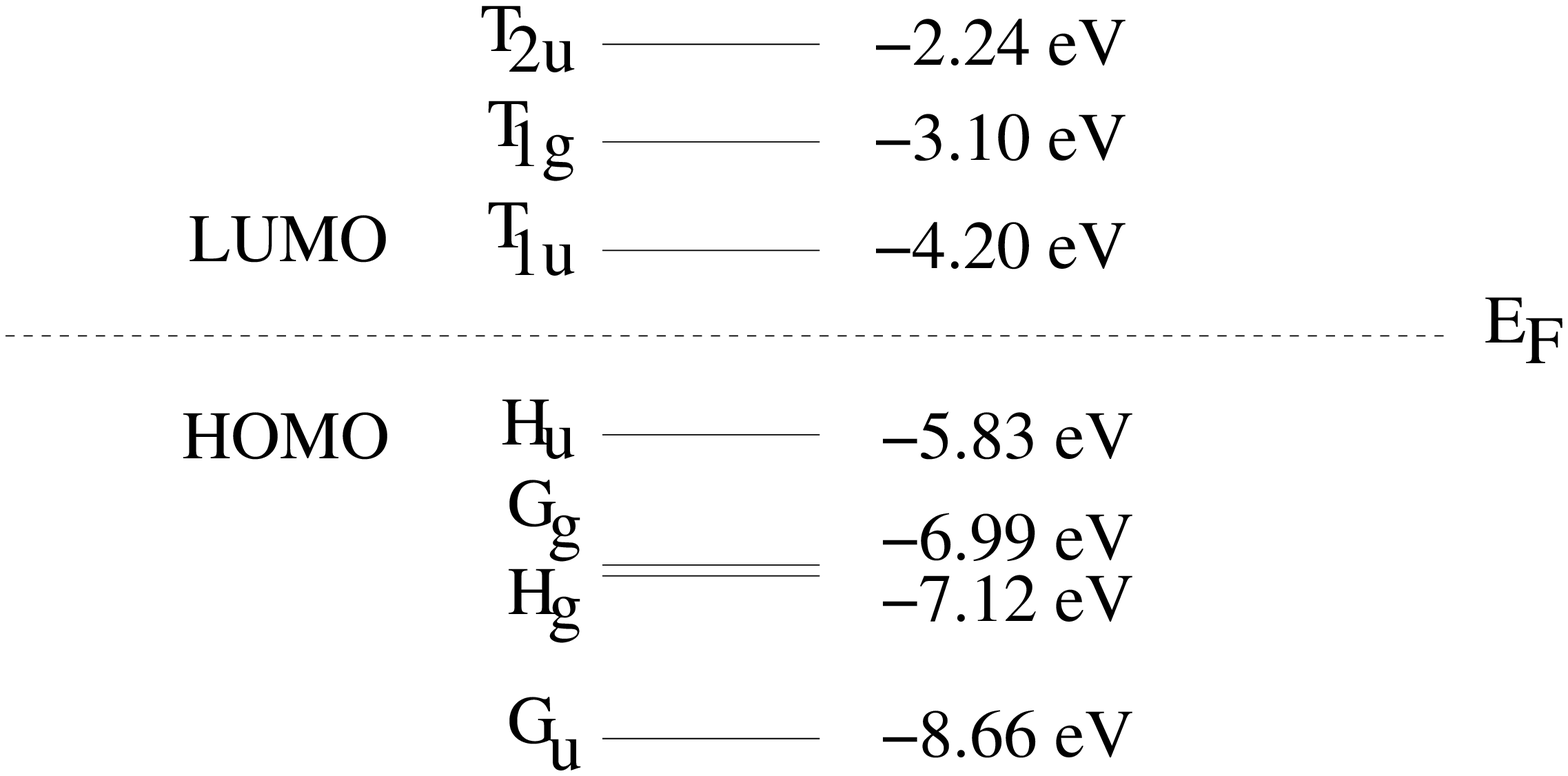}
  \end{center}
\end{minipage}\vspace{0cm}\\
\caption{\label{fig:c60Spectrum}
  Energy levels of the C$_{60}$ molecule as calculated by density functional
  theory. The Fermi energy for the undoped molecule is indicated. }
\end{center}
\end{figure}
%
In this section,
a minimal set of parameters describing the Stark effect in a neutral molecule 
is deduced using
the Wigner-Eckart theorem for the icosahedral
symmetry~\cite{book:Butler}. This is achieved in the framework of
density-functional perturbation theory~\cite{art:baroni1} where  
an external perturbation $V$ induces a change in the effective
potential (self-consistent field) $V\tu{eff}$
which couples the degenerate energy levels. In linear order the change 
is given by 
\begin{eqnarray}\label{eq:deltaVeff}
 \Delta V\tu{eff}(\VC r)=V(\VC r)+\Delta V\tu i(\VC r)
     +\frac{dv\tl{xc}}{dn}\Big|_{n=n(\VC r)}\Delta n(\VC r),
\end{eqnarray}
where 
$\Delta V\tu i(\VC r)=e^2\int d\VC r' \Delta n(\VC r')|\VC r - \VC r'|^{-1}$ 
is the
change in the induced potential due to the linear 
change $\Delta n(\VC r)$ in the
electron distribution. The last term is the exchange-correlation potential.
Within linear response, the change in the charge distribution $\Delta n$ 
as well as the
induced potential $\Delta V\tu i$ have the same symmetry (with respect to
$I_h$)  
as the external perturbation $V\tu e$. Furthermore, if $n(\VC r)$ is the
unperturbed charge distribution, then the factor 
$\frac{dv\tl{xc}}{dn}|_{n=n(\VC r)}$ in the last term 
of~(\ref{eq:deltaVeff}) has $A_g$ symmetry and does not change the symmetry of
$\Delta n$. Consequently, $\Delta V\tu{eff}$ has the same symmetry as 
$V\tu e$ and, using the notation of the previous section, can be written as 
\begin{equation} \label{eq:Veff}
  \Delta V\tu{eff}(\VC r)=\sum_{lxk}V_{lxk}\,f_{xk}^l(\VC r)+
      \MC O[V_{lxk}^2],
\end{equation}
where $f_{xk}^l(\VC r)$ are partner functions of the IR $x$ of $I_h$. Note
that $f_{xk}^l$ contains spherical harmonics 
$Y_{l'xk}$ with all $l'$ allowed by
\tab~\ref{tab:ldecompI}. The coupling of the degenerate levels is given by the
matrix elements of $\Delta V\tu{eff}$ with respect to the eigenstates of the
unperturbed molecule which are denoted by $\ket{nxk}$, where 
$n$ is the
quantum number differentiating between orbitals with the same IR $x$. 
In this context, the functions $f_{xk}^l$ play the role of
tensor operators of $I_h$ and the Wigner-Eckart theorem
can be used to write the matrix elements as  
\begin{eqnarray} \label{eq:cpl}
  \lefteqn{ \bra{n_2x_2k_2}\Delta V\tu{eff}\ket{n_1x_1k_1}= } \\
  & &{} \sum_{lxk}V_{lxk}\,
  \sum_{\lambda}t_{\lambda}(n_1x_1\,n_2x_2; lx)\,
  \MC{C}_{k_2k_1}^k(\lambda; x_2x_1;x).\nonumber
\end{eqnarray}
The coefficients $\MC C_{k_2k_1}^k(\lambda; x_2x_1;x)$ denote the 3jm symbols
(or Clebsch-Gordan coefficients)   
of $I_h$ and are entirely determined by the icosahedral symmetry.
In the present work they were taken
from Ref.~\onlinecite{book:Butler} and are discussed in detail in 
appendix~\ref{sec:coupling}.  
In order for $\MC C_{k_2k_1}^k(\lambda; x_2x_1;x)$ to be non-zero, the IR $x$
needs to be present in 
the decomposition of $x_1\otimes x_2$.
If $x$ occurs more than once in 
$x_1\otimes x_2$ then the multiplicity index $\lambda$ is required.
From this selection rule follows again that 
even potentials couple linearly whereas odd
potentials couple only in second order (quadratic Stark effect). 
Finally, the factors $t_{\lambda}(n_1x_1\,n_2x_2; lx)$ 
are the coupling constants which constitute the
minimal set of parameters describing the level splitting. 

We analyze in more detail
the level splitting of the LUMO ($x=T_{1u}$) and HOMO ($x=H_u$)
for $l=1$ and $l=2$ external potentials
which correspond to an electric field and a quadrupole potential. 
We will see in
Section~\ref{sec:mol_in_layer} that these two multipoles are
dominant in the case of a charged C$_{60}$ layer exposed to an electric field.
Within the icosahedral symmetry
$I_h$, the two potentials form partners for the IR $T_{1u}$ and $H_g$
respectively (see \tab~\ref{tab:ldecompI}). 
Details of the calculation
are given in appendix~\ref{sec:coupling}.  
The coupling constant
derived from the level splitting calculated by DFT 
are given in \tab~\ref{tab:cc}.
In the case of the odd $l=1$ potential, only second-order coupling to
closest-by orbitals was considered ($T_{1g}$ for the LUMO, $H_{g}$ and $G_{g}$
for the HOMO, see \fig~\ref{fig:c60Spectrum}), which, however, as shown
below, gives very satisfactory results.
For the splitting of the HOMO under the $l=2$ ($x=H_{g}$) potential, 
there are two
coupling constants because $H_{g}$ occurs twice in the product
$H_u\otimes H_u=A_g\oplus T_{1g}\oplus T_{2g}\oplus 2G_g \oplus 2H_g$.
In appendix~\ref{sec:coupling} the coupling matrix $H_{T_{1u}}'$ ($H_{H_u}'$) 
describing
the level splitting of the LUMO (HOMO), due to an applied 
$(lm)=(10)$ and $(lm)=(20)$ potential along an arbitrary direction of the
molecule, is calculated using perturbation theory. To a very good
approximation, the result can be cast into
the form
\begin{eqnarray}
  \label{eq:LUMOcpl0}
  H_{T_{1u}}'(\theta)&=&\Big(V_{10}^2\,c_1+V_{20}\,c_2\Big)\;
     \MC C_{T_{1u}}(\theta), \\
  \label{eq:HOMOcpl0}
  H_{H_u}'(\theta)&=&\Big(V_{10}^2\,d_1+V_{20}\,d_2\Big)\;
      \MC C_{H_u}(\theta),   
\end{eqnarray}
where $c_1$, $c_2$, $d_1$ and $d_2$ are constants depending on the coupling
constants of \tab~\ref{tab:cc} and the energies given in
\fig~\ref{fig:c60Spectrum}. $\MC C_{T_{1u}}(\theta)$ and  
$\MC C_{H_u}(\theta)$ 
are matrices within the LUMO and HOMO subspace respectively
and depend on the angle $\theta$ between the $z$-direction of the 
$(lm)=(10)$,~$(20)$ potentials and the 5-fold axis of the molecule. 
(for more details see appendix~\ref{sec:coupling}). 
In \fig~\ref{fig:LUMOSplitting} and \fig~\ref{fig:HOMOSplitting} 
the splitting of the LUMO and HOMO
is shown using the previous relations along with 
the points calculated by DFT. The group-theoretical fit is very satisfactory.
Note that the splitting of the
LUMO is independent of the orientation of the molecule when only
coupling among the LUMO or to the $T_{1g}$ is considered (see
appendix~\ref{sec:coupling}). Furthermore, relations~(\ref{eq:LUMOcpl0}) 
and~(\ref{eq:HOMOcpl0}) are quite remarkable, as they imply that the
contributions of the $l=1$ and $l=2$ potential lift the degeneracy of the
molecular levels in the same way, and thus the total splitting is given
by the sum of the $l=1$ and $l=2$ splittings.

%
\begin{table}
\begin{tabular}{cccc|c}
  $x_1$    & $x_2$    & $x$ & $\lambda$ & $t_{\lambda}(x_1\,x_2; x)$ \\
  \hline
  $T_{1u}$ & $T_{1g}$ & $T_{1u}$   & &    -0.663 \\
           & $T_{1u}$ & $H_g$      & &    -0.225 \\
  $H_{u}$  & $H_{g}$  & $T_{1u}$   & &    -0.730 \\
           & $G_{g}$  & $T_{1u}$   & &    -0.730 \\
           & $H_{u}$  & $H_g$      & 1 &  -0.520 \\
           & $H_{u}$  & $H_g$      & 2 &  -0.018 
\end{tabular} 
\caption{\label{tab:cc}
  Coupling constants $t_{\lambda}(x_1\,x_2; l)$ as defined by
  equation~(\ref{eq:cpl}) (indices $n_1$, $n_2$, $l$ are dropped)
  among the energy levels near the Fermi energy.  
  The coupling constants are given in units of 
  eV~$(7\,\bohr)^{l+1}e^{-1}$ where 
  $l=1$ for $x=T_{1u}$ and $l=2$ for $x=H_g$.
  The negative sign of the coefficients is due to the negative charge of the
  electrons.
  Note that the coupling constants $t(H_uH_g;T_{1u})$
  and $t(H_uG_g;T_{1u})$ are equal up to the third digit which is
  due 
  to fact that the $H_g$ and $G_g$ levels have almost the same radial
  dependence  because of the their closeness in energy (see
  Fig~\ref{fig:c60Spectrum}). 
} 
\end{table}
%
\begin{figure}
\begin{center}
\begin{minipage}[b]{0.4\textwidth}
  \begin{center}
    \includegraphics[width=\textwidth,clip=true]{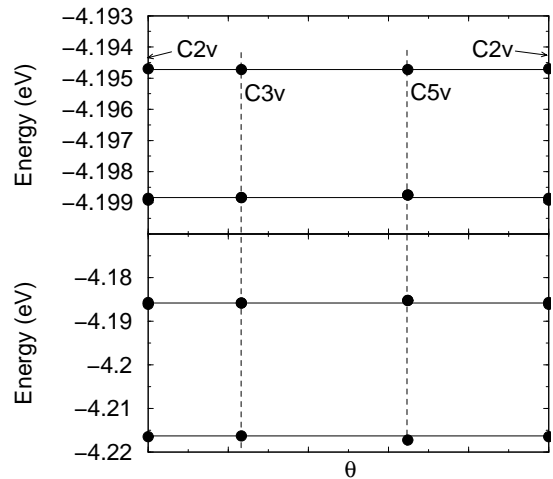}
  \end{center}
\end{minipage}\vspace{0cm}\\
\caption{\label{fig:LUMOSplitting}
  Splitting of the LUMO for an $(lm)=(10)$ and $(lm)=(20)$ potential as 
  the molecule is rotated by $\pi/2$ about the $y$-axis. The orientations
  shown in \fig~\ref{orient} are indicated. The splitting calculated with 
  DFT at these high symmetry orientations are indicated by the filled symbols.
  The lines give the result of perturbation theory, fitted to the
  calculation for the 3-fold axis. 
  \emph{Upper panel:} $(lm)=(10)$ potential with 
  $V_{10}=0.143\,e\,(7\,\bohr)^{-2}$. The lower level is twofold
  degenerate. 
  \emph{Lower panel:} $(lm)=(20)$ potential with 
  $V_{20}=0.180\,e\,(7\,\bohr)^{-3}$. 
  The upper level is twofold degenerate.  }
\end{center}
\end{figure}
%
\begin{figure}
\begin{center}
\begin{minipage}[b]{0.4\textwidth}
  \begin{center}
    \includegraphics[width=\textwidth,clip=true]{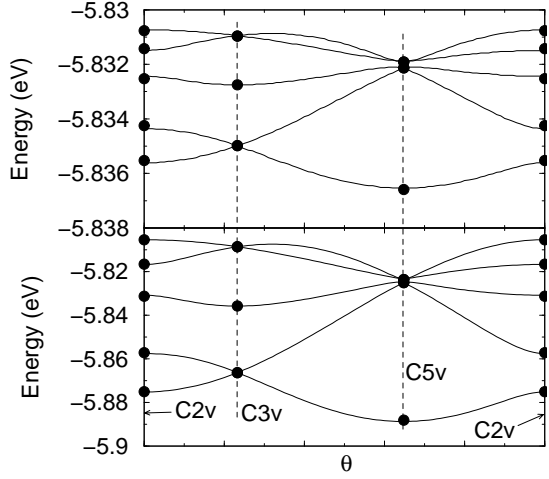}
  \end{center}
\end{minipage}\vspace{0cm}\\
\caption{\label{fig:HOMOSplitting}
 Splitting of the HOMO for an (lm)=(10) and (lm)=(20) potential.
 See \fig~\ref{fig:LUMOSplitting} for more details. In this plot the
 similarity of the splitting resulting from the different potentials
 (cf.~eqn.~(\ref{eq:HOMOcpl0})) is particularly striking.  }
\end{center}
\end{figure}

\section{Molecules in a Layer} \label{sec:mol_in_layer}
 
In this section a simple model of a C$_{60}$ FET is
considered. As mentioned above a FET can be understood as
capacitance where one plate is the gate and the other plate is the material
(here C$_{60}$) which is investigated. An analysis of this device, in
particular the calculation of the charge distribution, was done in a previous
work~\cite{art:wehrli1}. It was found, that the charge concentrates on the
first layer in the high doping regime. In what follows,  
we take this as a motivation to
consider a single layer of C$_{60}$ molecules which acts as a plate of a
capacitance. 
We use the response of a \emph{neutral} molecule (\tab~\ref{tab:lr}) to
describe the electrostatic behavior of the molecules in the layer. The doped
charge is taken care of by adding a monopole on every site.
The molecules are then exposed to the electric field arising from the gate as
well as the monopole fields of the neighboring molecules.  
In order to simplify the calculation, we assume a perfect lattice, either
square, for the (001), or triangular, for the (111) plane of 
the fcc lattice formed by the C$_{60}$ molecules in the bulk.
In order for the sites to be
equivalent we neglect 
the non-spherically symmetric part of the response of the C$_{60}$-molecule
and use the averaged response given by
\begin{equation}\label{eq:ssalpha}
  \alpha_{l_1m_1\,l_2m_2}\tu{spher.-sym.}=\delta_{l_1l_2}\,
           \delta_{m_1m_2}\,\sum_{x\,k}\frac{\alpha_{l_1xk\,l_1xk}}{2l_1+1},
\end{equation}
where the sum is taken over all components of the $l_1$ subspace.
Since the response of C$_{60}$ for multipoles $l\le2$ is isotropic
(cf.\ section \ref{sec:irresp}), this averaging is exact for $l=1,2$.

\begin{figure}
\begin{center}
\begin{minipage}[b]{0.35\textwidth}
  \begin{center}
    \includegraphics[width=\textwidth,clip=true]{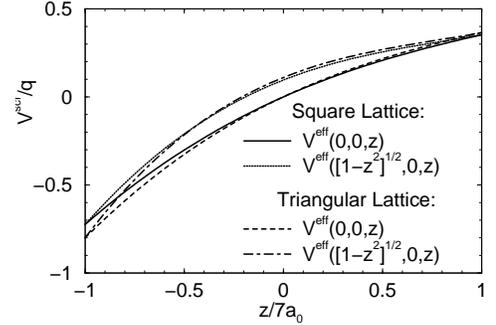}
  \end{center}
\end{minipage}\vspace{0cm}\\
\caption{\label{fig:ScrPot} 
  Screened potential $V\tu{scr}$ across the layer with the parameters from
  Tab~\ref{tab:ScrPot} for the square and triangular lattice. 
  The gate is assumed to be on the left. The lower lines
  correspond to a cut through the center of the 
  molecule whereas the upper lines
  are along half-circles with the radius of the molecule
  ($7\,\bohr\approx3.7$~\AA{}).  
}
\end{center}
\end{figure}

Because of translational invariance, the total potential is given by
\begin{equation}
  V\tl{tot}(\VC r)=V\tu e(\VC r) +\sum_{\VC R_i}V\tu i(\VC r-\VC R_i) ,
\end{equation}
where the sum is taken over all lattice sites $\VC R_i$. 
At a given site, say $\VC R_i=0$, the
total potential can also be decomposed as 
$V\tl{tot}(\VC r)=V\tu{scr}(\VC r) +V\tu i(\VC r)$ 
where the screened potential 
$V\tu{scr}(\VC r)=V\tu e(\VC r) +\sum_{\VC R_i\ne 0}V\tu i(\VC r-\VC R_i)$
contains the external potential $V\tu e(\VC r)$ as well as the 
sum of all induced potentials of the other molecules. This sum    
depends linearly on the induced potential $V\tu i$ and, hence, the
coefficients of $V\tu{scr}$ are given by 
\begin{equation}\label{eq:vscr1}
  V\tu{scr}_{l_1m_1}=V_{l_1m_1}+
    \sum_{l_2m_2}\,\beta_{l_1m_1\,l_2m_2} Q_{l_2m_2}.
\end{equation}
The matrix $\beta$ is entirely given by geometry and discussed in
appendix~\ref{sec:shtranslation}.  
On the other hand, the screened potential $V\tu{scr}$ induces
a potential $\Delta V\tu i$ as given in equation~(\ref{eq:lr}). 
Therefore the total induced potential $V\tu i=V\tu{i,0}+\Delta V\tu i$ is 
\begin{equation}\label{eq:vscr2}
  Q_{l_1m_1}=Q^0_{l_1m_1}-
    \sum_{l_2m_2}\,\alpha_{l_1m_1\,l_2m_2} V\tu{scr}_{l_2m_2}.
\end{equation}
Equations~(\ref{eq:vscr1}) and~(\ref{eq:vscr2}) 
can be combined by eliminating the
coefficients $Q_{lm}$ which yields
\begin{eqnarray}\label{eq:vscr3}
  V\tu{scr}_{l_1m_1}&=&\sum_{l_2m_2}[1+\beta\alpha]^{-1}_{l_1m_1\,l_2m_2}
    V\tu{bare}_{l_2m_2}, \\ \label{eq:vbare}
  V\tu{bare}_{l_1m_1}&=&V_{l_1m_1}+
    \sum_{l_2m_2}\beta_{l_1m_1\,l_2m_2}Q^0_{l_2m_2},
\end{eqnarray}
where the $V\tu{bare}_{l_1m_1}$ describes the bare potential arising from the
external potential (the electric field of the gate) and intrinsic moments of
the molecules (the induced charge, i.e. monopoles).   
The square (triangular) lattice in the presence of the electric field has
the rotational symmetry $C_{4v}$ ($C_{6v}$).
As a consequence, only components with these symmetries are
non-zero and therefore they are given by $\Re(Y_{lm})$ 
with m a multiple of 4 (6).
Using relation~(\ref{eq:vscr3}) and~(\ref{eq:vbare}) 
the screened potential can be calculated. 
The non-zero components entering~(\ref{eq:vbare}) are the monopole charge
$Q^{0}_{00}=q$ and electric field $V_{10}=-E\tl{Gate}$. As 
the FET is overall neutral
$E\tl{Gate}=-2\pi q/A\tl{mol}$ with 
$A_{(001)}=a^2/2$ for the square lattice and 
$A_{(111)}=\sqrt{3}a^2/4$ for the triangular lattice.
The results are given in \tab~\ref{tab:ScrPot} and
graphically depicted in \fig~\ref{fig:ScrPot}.
The components $(lm)=(10),\,(20)$ 
are most dominant and higher ones are at least one
order of magnitude smaller. This justifies a posteriori the 
assumption of spherical
symmetry because~(\ref{eq:ssalpha}) is exact for $(lm)=(10),\,(20)$.
From \fig~\ref{fig:ScrPot} it can be seen that the electric field is
efficiently screened within the layer. Note that decrease of the electric
field yields negative sign of $v_{20}$ in \tab~\ref{tab:ScrPot}. 
We also have checked the influence of adjacent layers of C$_{60}$ and of a
close-by dielectric (with a dielectric constant $\varepsilon=10$). The effects
on the parameters in \tab~\ref{tab:ScrPot} were less than 2~\%. The reason
is that the field inhomogeneities induced by a 2D lattice of multipoles
decay exponentially outside the lattice.
%
\begin{table}
\begin{tabular}{c|D{.}{.}{4}|D{.}{.}{4}||D{.}{.}{4}|D{.}{.}{4}}
       & \multicolumn{2}{c||}{square} & \multicolumn{2}{c}{triangular}  \\
  $lm$ & \multicolumn{1}{c|}{$v_{lm}\tu{bare}$}
       & \multicolumn{1}{c||}{$v_{lm}\tu{scr}$}
       & \multicolumn{1}{c|}{$v_{lm}\tu{bare}$}
       & \multicolumn{1}{c}{$v_{lm}\tu{scr}$} \\
  \hline
  $10$ & 0.862   & 0.499  & 0.996  & 0.530 \\
  $20$ & -0.230  & -0.189 & -0.280 & -0.219\\
  $30$ & 0       & 0.039  & 0      & 0.053\\
  $40$ & 0.0133  & 0.0030 & 0.0177 & 0.0015 \\
  $4c4$& 0.0174  & 0.0097 & \multicolumn{1}{c|}{-} & \multicolumn{1}{c}{-} 
  \end{tabular} 
\caption{\label{tab:ScrPot}
  Components of the bare and screened potential for the   
  square and triangular lattice with $a/\sqrt{2}=10$~\AA.
  The coefficients are in units of $q/(7\,\bohr)^{l+1}$, 
  where $q$ is the charge per
  C$_{60}$ molecule. Note that $7\,\bohr\approx3.7$~\AA{}
  is about the radius of the C$_{60}$ molecule. } 
\end{table}
%

\section{Splitting in self-consistent multipole field}\label{sec:scsplit}

\begin{figure}
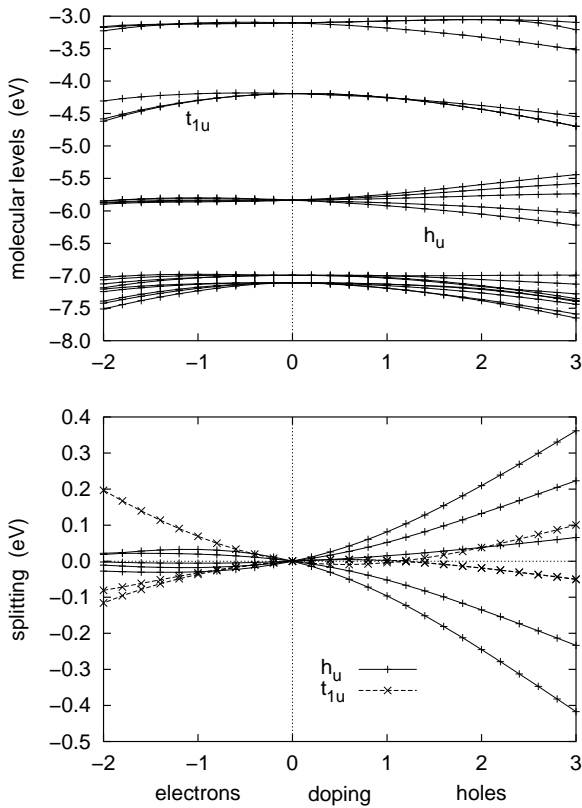

 \centering
 \resizebox{3in}{!}{\rotatebox{270}{\includegraphics{C2vveffsqul.epsi}}}\\
 \resizebox{3in}{!}{\rotatebox{270}{\includegraphics{C2vveffsqus.epsi}}}
 \caption[]{\label{scstark}
            Splitting of the molecular levels of a C$_{60}$ molecule in 
            the self-consistent multipole potential ($l\le2$) for a (001)
            monolayer (square lattice) in a homogeneous external field
            as a function of the induced charge. The molecule is oriented
            such that one of its two-fold axis points in the direction of
            the external field (perpendicular to the monolayer). The top 
            panel shows the positions of the split (from bottom to top)
            $H_g$, $G_g$, $H_u$, $T_{1u}$, and $T_{1g}$ levels. The bottom 
            gives the splitting of the $H_u$ (HOMO) and $T_{1u}$ (LUMO) levels
            relative to their respective center of gravity.}
\end{figure}

\begin{figure}
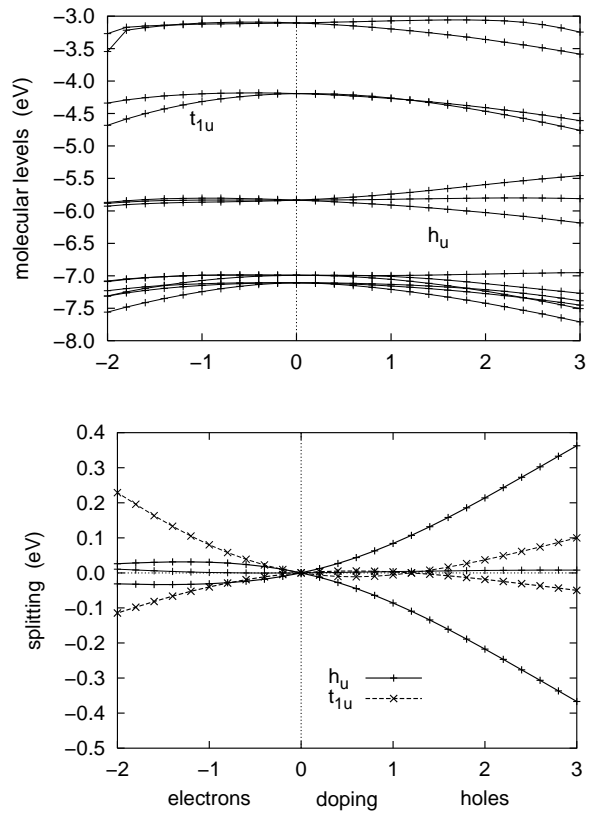

 \centering
 \resizebox{3in}{!}{\rotatebox{270}{\includegraphics{C3vvefftril.epsi}}}\\
 \resizebox{3in}{!}{\rotatebox{270}{\includegraphics{C3vvefftris.epsi}}}
 \caption[]{\label{scstarkC3v}
            Splitting of the molecular levels of a C$_{60}$ molecule in
            the self-consistent multipole potential ($l\le2$) for a (111)
            monolayer (triangular lattice) in a homogeneous external field
            as a function of the induced charge. The molecule is oriented
            such that one of its three-fold axis points in the direction of
            the external field (perpendicular to the monolayer). Otherwise
            the plots are as in figure \ref{scstark}. The thick lines in
            the lower panel indicate the two-fold degenerate levels.}
\end{figure}

We are now in the position to estimate the effect of the external field
on the electronic structure of the C$_{60}$ molecules in the monolayer that
carries charge. To do so we have performed density functional calculations 
for a molecule in the self-consistent multipole fields as determined in the
previous section (cf.~\tab~\ref{tab:ScrPot}).
\Fig~\ref{scstark} shows the splitting of the molecular levels in the
self-consistent field for a (001) monolayer, where the molecule is oriented
with one of its two-fold axes pointing in the direction of the external field.
The maximum energy difference between split states is given in
\tab~\ref{tab:MaxSplit} and compared to the result from the perturbative
formula~(\ref{eq:LUMOcpl0}) and~(\ref{eq:HOMOcpl0}), which is in good
agreement for $|q|\le 2$. 
As \fig~\ref{scstarkC3v} demonstrates, similar results are obtained for other
geometries. Qualitatively, the results are also in agreement with an 
approximative
tight-binding calculation published earlier in Ref.~\onlinecite{art:swehrli02}.
As expected we find that the stronger the external field, i.e., 
the larger the induced charge, the stronger the splitting. We notice also
a pronounced asymmetry in the splitting: when the monolayer is charged with
electrons the splitting is different from when it is charged with holes.
Again, the reason is parity: Because of parity an external homogeneous 
field, and more generally any multipole potential with $l$ odd, leads to a 
quadratic Stark effect. Hence, for odd $l$ the splitting is independent of 
the sign of the field. For $l$ even, however, the levels split already in first 
order (linear Stark effect), so the splitting changes sign with the external 
field. Thus the asymmetry originates from the multipoles with even $l$. 
Moreover, because of the first-order versus second-order
effect, even though the largest even multipole (20) is significantly smaller 
than the largest odd multipole (the screened external field), it contributes 
considerably to the splitting.

We have seen in section \ref{sec:split} that for the
HOMO and LUMO of C$_{60}$ the splittings caused by (10) and (20) multipole
potentials are essentially additive (cf.~equation~(\ref{eq:cpl})). 
Thus we expect them to add up for one sign of the external field, while
they should compensate for the opposite sign. In fact, for electron doping
the splittings of the HOMO seem to almost perfectly cancel, while upon
hole doping the splitting of the HOMO is essentially doubled compared to
the splitting caused by the screened homogeneous field alone. Thus for
the HOMO, the splittings happen to add up for an external field that 
induces charge carriers into that orbital --- a situation that is
particularly unfavorable for hole doping. For the LUMO the situation
is similar: For an external field that induces electrons in the LUMO
the splitting is enhanced. So it turns out that the contributions 
of the higher multipoles conspire to enhance the splitting of the orbital
that carries the induced charge. For both, HOMO and LUMO, the splitting
becomes comparable to the respective band width for a field
that corresponds to about two charge carriers per molecule.

The calculations reported above have been performed for an uncharged
molecule. Considering instead the splitting for a molecule that carries
the proper induced charge, the splittings are substantially reduced. This is
easily understood: Due to the splitting, the electrons will fill only
the lowest of the split levels. But since the interaction between electrons
in the same orbital is larger than the interaction of electrons in different
orbitals, the occupied levels will be shifted upwards, compared
to the levels that were left empty -- thereby reducing the splitting.
We are, however, not interested in the splitting per se, but in the effect
of the splitting on the band structure of a monolayer. Thus allowing electrons
only in the energetically lowest levels would mean that in the lattice,
the electrons are not allowed to hop to energetically higher ones of the split
levels.  This implies that the original band structure would already be 
separated into a set of bands originating from occupied, and another set of 
bands originating from the empty orbitals.
To eliminate the undesired differences in the interaction between
electrons in the split orbitals, we therefore work with the splittings 
obtained for a neutral molecule.


\begin{table}
\begin{tabular}{c||c|c||c|c}
       & \multicolumn{2}{c||}{LUMO} &  \multicolumn{2}{c}{HOMO} \\
  $q$  & DFT & Pert. &   DFT & Pert. \\
  \hline
   -2e  & 0.312 & 0.305 &  0.049 & 0.001\\
   -1e  & 0.105 & 0.102 &   0.062 & 0.060 \\
   0    & 0 &  0 & 0 & 0 \\
   1e   & 0.006 & 0.002 & 0.178&  0.179 \\
   2e   & 0.057 & 0.097 & 0.455 & 0.476  \\
   3e   & 0.151 & 0.297 & 0.779 & 0.892 \\
\end{tabular} 
\caption{\label{tab:MaxSplit}
  Maximum energy difference (in eV) between split HOMO and LUMO 
  states, respectively, as a function of doping $q$ of the square lattice.
  2nd and 4th column are the DFT results from \fig~\ref{scstark}. 
  3rd and 5th column are calculated by perturbation
  theory as described in section~\ref{sec:split}.  }
\end{table}

\begin{figure}
\begin{center}
\begin{minipage}[b]{0.35\textwidth}
  \begin{center}
    \includegraphics[width=\textwidth,clip=true]{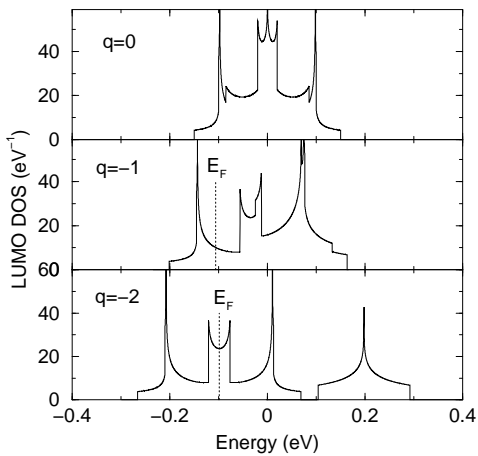}
  \end{center}
\end{minipage}\vspace{0cm}\\
\caption{\label{fig:LUMODOS} 
  DOS (per molecule and for both spins) 
  of the LUMO-band taking into account the level splitting for doping
  $q=0,-1,-2$. The Fermi energy is indicated.  }
\end{center}
\end{figure}

\begin{figure}
\begin{center}
\begin{minipage}[b]{0.35\textwidth}
  \begin{center}
    \includegraphics[width=\textwidth,clip=true]{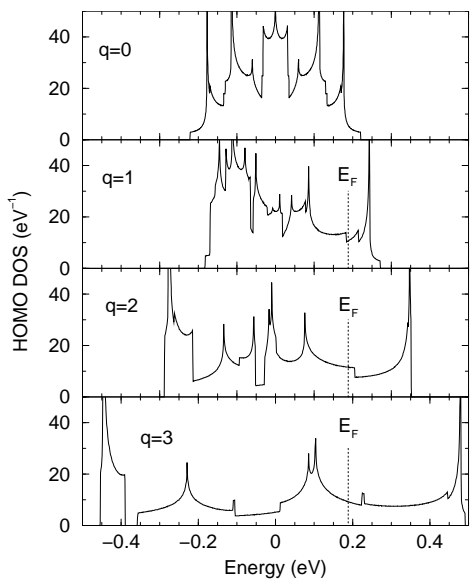}
  \end{center}
\end{minipage}\vspace{0cm}\\
\caption{\label{fig:HOMODOS} 
  As in \fig~\ref{fig:LUMODOS} but for the DOS of the HOMO-band with
  $q=0,1,2,3$ .
}
\end{center}
\end{figure}

To estimate the effect of the Stark splitting on the density of states (DOS),
we have performed tight-binding calculations for the (001) monolayer (square 
lattice), assuming the unidirectional structure (two-fold axis of the molecules
pointing in the direction of the external field). 
The basis for the tight-binding Hamiltonian and the hopping matrix elements
were taken from the parametrization given in 
Ref.~\onlinecite{art:satpathy92}.
The splittings
shown in \fig~\ref{scstark} were then used to derive an on-site coupling
between the different orbitals. In the case of the LUMO, the on-site coupling
is diagonal and reduces to orbital dependent on-site energies.
The DOS for the LUMO ($T_{1u}$) and the HOMO ($H_u$) bands calculated with this 
model for different charging are shown in \fig~\ref{fig:LUMODOS} and 
\ref{fig:HOMODOS}, respectively. We find that already for an induced charge
of one carrier per molecule the change with respect to the unperturbed DOS
is sizable. For $q=-2$ one of the $T_{1u}$-bands is already completely
separated from the other two. Also the HOMO density of states shows for $q=2$ 
hardly any resemblance to the original DOS, and for $q=3$ also the $H_u$-bands 
fall into two groups. 
We thus conclude that beyond filling $|q|=2$ the electronic structure 
is distorted so much compared to the unperturbed monolayer that one can
no longer speak of doping.

The calculation of the density of states in a minimal tight-binding 
basis involves, of course, approximations:
First, in the lattice, not only hopping between LUMO (HOMO) levels 
is allowed, but also hopping via energetically 
close-by levels. An orbital at $\Delta\varepsilon$
away will give a contribution to the hopping of about $t^2/\Delta\varepsilon$,
where $t$ is the hopping matrix element from the orbital, that we consider 
explicitly, to the orbital at $\Delta\varepsilon$. The influence of this
effect on the hopping between molecules was studied in
Ref.~\onlinecite{art:satpathy92} and changes of the order of 5~\% were
found. 
More importantly, due to the deformation of the molecular orbitals in the
field, the hopping matrix elements between the HOMO or LUMO orbitals
will change. For a simple estimate, we have performed tight-binding
calculations of a C$_{60}$ molecule in an external homogeneous field
and determined the average hopping matrix element between the 
$t_{1u}$-orbitals following the approach of Ref.~\onlinecite{hopp}.
We find that the change in the hopping matrix elements depends strongly
on the orientation of the molecules. Typical changes are of the order of
10--20\%.

%

\section{Conclusions}\label{sec:concl}

We have analyzed the changes in the electronic structure of a C$_{60}$ 
monolayer in which charge carriers are induced by the application of an
external homogeneous electric field. 
We find that the effective field seen by each molecule in the monolayer
is strongly screened, but that there are additional higher multipole 
potentials. Although these components are considerably weaker than the
screened homogeneous field, for even $l$, they give a significant 
contribution to the level splitting as they are of first order.
In addition the $l=1$ (homogeneous field) and $l=2$ potentials split the
HOMO and LUMO in almost the same way, so the splittings they produce add up 
or counteract, depending on the sign of the external field.
For both, the HOMO and the LUMO, the signs turn out to be such, 
that the splitting is enhanced when the charge is induced in the respective
level. Thus the level that carries the field-induced charge is strongly
changed by the effective field --- a particularly unfavorable situation if
one wants to achieve doping, i.e., filling of a level without substantially
changing its electronic structure.

There are, of course, some effects that have been neglected in our analysis:
The polarizability of C$_{60}$ increases when charge is put in the
$T_{1u}$ orbital. Thus when filling the LUMO the screening of the external
field should become somewhat more efficient. Second, due to the electric field 
the molecular orbitals are deformed and thus the hopping matrix elements 
between neighboring molecules change. This effect depends sensitively on
the orientation of the molecules but it typically leads to a slight decrease 
of the band width with applied field, making the Stark splitting even more
important. Furthermore we have neglected the effect of electron-phonon coupling,
which also tends to narrow the bands. This should be particularly important for
the $H_u$-band.\cite{elph}
Finally, we have not considered correlation effects. It is known that,
at integer filling, the alkali-doped Fullerenes are close to a Mott transition
and that they are metallic because of orbital degeneracy.\cite{c60mott} 
Due to the reduced coordination in two dimensions, a doped monolayer of C$_{60}$
should be even more strongly correlated, even if the orbitals are still
degenerate. Lifting the degeneracy, e.g., by the Stark effect significantly 
increases correlations even further.\cite{manini} It therefore seems inevitable
that for an integer number of induced charges a C$_{60}$ monolayer will
be a Mott insulator, even though it is not clear what effect the gate oxide
might have on the Coulomb repulsion $U$ between two electrons on a C$_{60}$
molecule.\cite{kirchb03} 

To summarize, it seems safe to conclude that the physics of field-effect 
devices based on C$_{60}$ as active material should be quite different 
from that of the alkali-doped fullerenes, at least at doping levels beyond
two carriers per molecule.

\begin{acknowledgments}
  The authors would like to thank O.~Gunnarsson, T.M.~Rice, B.~Batlogg, and
  C.~Helm for fruitful discussions. This work has been supported by the Swiss
  Nationalfonds.
\end{acknowledgments}

\appendix

\section{Response for general orientation} \label{sec:rsh}

To illustrate the use of the response coefficients given in \tab~\ref{tab:lr},
we show, how to calculate the response for a multipole
potential with $(lm)=(30)$ for different orientations of the $z$-axis with
respect to the C$_{60}$ molecule. This is the first non-trivial case, as the
response for multipoles with $l\le2$ is isotropic.
To start with, we need the basis functions $Y_{lxk}$ spanning the irreducible 
representation of the $I_h$, that were introduced in section \ref{sec:irresp}.
For specific orientations, these can be found, e.g., in the Ref.\ 
\onlinecite{book:Butler}, chapter 16, or Ref.\ \onlinecite{dresselhaus}, 
\tab~4.2. 
For arbitrary orientations, they have to be derived by explicitly finding 
the basis functions that span the irreducible representations of the
icosahedral group. For the sake of the example, we consider
the response for $z$ along the 5-fold and the 3-fold axis of the C$_{60}$
molecule. The corresponding basis functions are then both found in Ref.\ 
\onlinecite{book:Butler}. Since we are interested in the response to an
external multipole with $(lm)=(30)$, we have to identify those basis functions
that contain $Y_{30}$. They are shown in \tab~\ref{tab:l3sh}.
It turns out that for $z$ parallel to the 5-fold axis the $(lm)=(30)$ potential
corresponds to a pure $T_{2u}$ potential. Thus the response is given by
$\Delta Q_{30}=\alpha_{33}(T_{2u})\,V_{30}$.
For $z$ along the 3-fold axis the situation is more complicated, as the
potential is now a mixture of partner functions of $G_u$ and $T_{2u}$:
As can be seen from \tab~\ref{tab:l3sh}, it mixes with the 
$Y_{3c3}=\sqrt 2\,\Re(Y_{33})$ component.
By construction, the response matrix $\alpha$ in the subspace
spanned by $\{Y_{30},Y_{3c3}\}$ is diagonal, with diagonal elements
$\alpha_{33}(H_u)$ and $\alpha_{33}(T_{2u})$, when written in the basis
functions listed in \tab~\ref{tab:l3sh}. However, to obtain the multipole 
response we have to use the basis $(Y_{30},Y_{3c3})$, for which there
are off-diagonal elements:

\begin{table}
\begin{tabular}{c | c | c}
  IR of $I_h$  & 5-fold axis &  3-fold axis \\
  \hline
  $G_u$    & -& $\frac{2\sqrt 2}{3}Y_{30}-\frac{1}{3}Y_{3c3}$  \\
  $T_{2u}$ & $-Y_{30}$ & $-\frac{1}{3}Y_{30}-\frac{2\sqrt 2}{3}Y_{3c3}$ 
\end{tabular} 
\caption{\label{tab:l3sh}
  Transformed $l=3$ spherical harmonics, which are partner functions for the IR
  of $I_h$ and which contain $Y_{30}$. The second (third) column is for the
  case where the 3-fold (5-fold) axis of the molecule is parallel to the 
  $z$-axis. The real spherical harmonic is $Y_{3c3}=\sqrt2\,\Re(Y_{33})$.}
\end{table}
\begin{widetext}
\begin{equation}
  \alpha = \frac{1}{9}
  \left( \begin{array}{cc}
      8\,\alpha_{33}(H_u)+ \alpha_{33}(T_{2u}) & 
      \sqrt 8\,[\alpha_{33}(T_{2u})-\alpha_{33}(H_u)] \\
      \sqrt 8\,[\alpha_{33}(T_{2u})-\alpha_{33}(H_u)]  & 
      \alpha_{33}(H_u)+ 8\,\alpha_{33}(T_{2u})
    \end{array} \right) .
\end{equation}
We thus find
\begin{eqnarray*}
  \Delta Q_{30}&=&\frac{1}{9}\left[
  8\,\alpha_{33}(H_u)+ \alpha_{33}(T_{2u})\right]\,V_{30},\\
  \Delta Q_{3c3}&=& \frac{\sqrt 8}{9}\,
    [\alpha_{33}(T_{2u})-\alpha_{33}(H_u)]\,V_{30}.
\end{eqnarray*}

\section{Coupling Matrices} \label{sec:coupling}

In this section we discuss the calculation of the level splitting for arbitrary
directions within
perturbation theory using the
coupling constants of \tab~\ref{tab:cc}. 
As discussed above, we restrict the analysis  
to $l=1$ and $l=2$ external potentials, which
corresponds to $x=T_{1u}$ and $x=H_g$ potentials in the icosahedral symmetry
$I_h$. In first order perturbation theory, 
the splitting of the levels in the degenerate
subspace $\MC E_{nx}$ is given by the
eigenvalues of the matrix
\begin{equation} \label{eq:firstorder}
  H_{k_1k_2}^{(1)}(nx)=\bra{nxk_2}V\tu{eff}\ket{nxk_1}.
\end{equation}
This matrix vanishes in the case of an odd potential and the
splitting is given by the second order expression
\begin{equation} \label{eq:secondorder}
  H_{k_1k_2}^{(2)}(nx)=\sum_{(n'x')\atop \ne (nx)}\sum_{k'}
      \frac{\bra{nxk_2}V\tu{eff}\ket{n'x'k'}\bra{n'x'k'}V\tu{eff}\ket{nxk_1}}
           {E_{nx}-E_{n'x'}}.
\end{equation}
The matrix-elements in~(\ref{eq:firstorder}) and~(\ref{eq:secondorder}) are
given in~(\ref{eq:cpl}) and involve the icosahedral Clebsch-Gordan coefficients
$\MC{C}_{k_2k_1}^k(\lambda; x_2x_1;x)$. In order for the $k$-indices to be
defined we consider the molecule oriented with the 5-fold axis
parallel to the $z$-axis (see \fig~\ref{orient}). This allows to label the
states within a multiplet unambiguously with its $C_5$ index $k$.
The ordered basis of a $T_{1u}$ subspace has $k$-indices 
$(0,1,-1)$ whereas the ordered basis of an $H_u$ subspace is given
by $(0,1,-1,2,-2)$. Note that in the case of applied $l=1$ or $l=2$ potential,
the $k$ index corresponds to the $m$ index of the spherical harmonics. 
For a detailed discussion we refer to Ref.~\onlinecite{book:Butler}.
We will present the coefficients 
$\MC{C}_{k_2k_1}^k(\lambda; x_2x_1;x)$ as matrices with respect to the indices
$k_1$ and $k_2$.
In order to reduce the number of matrices, we will give the coupling matrices
for $(lm)=(10),(20)$ potentials, which are rotated around the $y$-axis
by an angle $\theta$. 
The resulting matrices are then given by
\begin{equation}
  \MC{C}^{\theta}(\lambda; x_2x_1;x)=
  \sum_k\,R_{0k}^{l}(\theta)\, \MC{C}^k(\lambda; x_2x_1;x),
\end{equation} 
where $R_{k'k}^{l}(\theta)$ is the rotation matrix of the 
spherical harmonics in a given $l$-subspace. 
Using the previous relations and equation~(\ref{eq:cpl}), the   
coupling matrix for an even $l=2$ potential is given by
\begin{equation} \label{eq:firstorderapp}
  H^{(1)}(nx,\theta)=V_{20}
  \sum_{\lambda}\,t_{\lambda}(nx\,nx;H_g) \,\MC C_{\lambda}^{\theta}(xx;H_g).
\end{equation}
Note that the multiplicity label $\lambda$ is only relevant for the HOMO $H_u$
because $H_g$ occurs twice in the product 
$H_u\otimes H_u=A_g\oplus T_{1g}\oplus T_{2g}\oplus 2G_g \oplus 2H_g$.
The coupling matrix for the odd $l=1$ potential is
\begin{equation} \label{eq:secondorderapp}
  H^{(2)}(nx,\theta)=V_{10}^2
  \sum_{(n'x')\atop \ne (nx)}\frac{t(nx\,nx';T_{1u})^2}{E_{nx}-E_{n'x'}}\,
    \MC C^{\theta}(xx';T_{1u})^{\textrm{T}}\, \MC C^{\theta}(xx';T_{1u}).
\end{equation}
In the following we restrict the sum over 
subspaces $\MC E_{n'x'}$ closest in energy to 
$\MC E_{nx}$ which is the  $T_{1g}$ subspace in the case of the LUMO and 
the $H_{g}$ and $G_{g}$ subspaces in the case of the the HOMO 
(see Fig~\ref{fig:c60Spectrum}). Below, the coupling matrices 
$\MC C_{\lambda}^{\theta}(x_1x_2;x)$ which are needed to calculate the
splitting of the HOMO and LUMO are given. 
They are traceless $\textrm{Tr}\,\MC C^{\theta}=0$ and normalized such
that $\textrm{Tr}\,\MC (C^{\theta})^{\textrm{T}} \MC C^{\theta}=1$. 
The coupling matrices which describe the splitting of the LUMO are
\begin{eqnarray}
  \label{eq:l1t1ucpl}
  \MC C^{\theta}(T_{1u}T_{1g};T_{1u})&=&
  \frac{1}{2}
  \left( \begin{array}{ccc}
      0 & 
      -\sin \theta & 
      -\sin \theta \\
      -\sin \theta & 
      -\sqrt 2 \cos \theta & 
      0 \\ 
      -\sin \theta & 
      0 & 
      \sqrt 2 \cos \theta  
    \end{array} \right), \\
  \label{eq:l2t1ucpl}
  \MC C^{\theta}(T_{1u}T_{1u};H_g)&=&
  \frac{1}{4}
  \left( \begin{array}{ccc}
     -4\ {\sqrt{\frac{2}{3}}} + 2\ {\sqrt{6}}\ \sin^2 \theta  & 
     \sqrt{3}\sin 2\theta & 
     -\sqrt{3} \sin 2\theta  \\
     \sqrt{3} \sin 2\theta & 
     2\sqrt{\frac{2}{3}} - \sqrt{6} \sin^2 \theta  & 
     \sqrt{6} \sin^2 \theta  \\
     -\sqrt{3} \sin 2\theta   & 
     \sqrt{6} \sin^2 \theta  & 
     2\sqrt{\frac{2}{3}} - \sqrt{6} \sin^2 \theta
    \end{array} \right). 
\end{eqnarray}
The eigenvalues of these two matrices are independent of $\theta$
and given by $(0,-\frac{1}{\sqrt 2},\frac{1}{\sqrt 2})$ and 
$(-\frac{2}{\sqrt 6},\frac{1}{\sqrt 6},\frac{1}{\sqrt 6})$
which implies that the splitting is independent of the
orientation of the molecule with respect to the direction of the applied 
$l=1$ and $l=2$ potentials.
The coupling of the HOMO ($H_u$) to the lower lying $H_g$ and $G_g$ levels is
given by
\begin{eqnarray}
  \label{eq:l1hgcpl}
  \MC C^{\theta}(H_u H_g;T_{1u})&=&
  \frac{1}{2\sqrt 5}
  \left( \begin{array}{ccccc}
    0&
    \sqrt 3\sin \theta&
    -\sqrt 3\sin \theta&
    0&
    0\\
    \sqrt 3\sin \theta&
    -\sqrt 2 \cos\theta &
    0&
    \sqrt 2 \sin\theta &
    0\\
    -\sqrt 3\sin \theta&
    0&
    \sqrt 2 \cos\theta &
    0&
    -\sqrt 2 \sin\theta \\
    0&
    \sqrt 2 \sin\theta &
    0&
    - 2\sqrt 2 \cos\theta&
    0\\
    0&
    0&
    -\sqrt 2 \sin\theta
    &0
    &2\sqrt 2 \cos\theta
  \end{array} \right), \\
  \label{eq:l1ggcpl}
  \MC C^{\theta}(H_u G_g;T_{1u})&=&
  \frac{1}{2\sqrt 5}
  \left( \begin{array}{ccccc}
      \sqrt 3\sin \theta&
      2\sqrt 2 \cos\theta&
      0&
      -\frac{1}{\sqrt 2}\sin \theta&
      0\\
      -\sqrt 3\sin \theta&
      0&
      - 2\sqrt 2 \cos\theta&
      0&
      \frac{1}{\sqrt 2}\sin \theta\\
      0&
      -\sqrt 2\sin \theta\ &
      0&
      - \sqrt 2 \cos\theta&
      -\frac{3}{\sqrt 2}\sin \theta\\
      0&
      0&
      \sqrt 2\sin \theta&
      \frac{3}{\sqrt 2}\sin \theta&
      \sqrt 2 \cos\theta
  \end{array} \right). 
\end{eqnarray}
The coupling of the HOMO among themselves is given by 
\begin{eqnarray}
  \label{eq:l2hu1cpl}
  \MC C^{\theta}(1;H_u H_u;H_g)&=&
  \frac{1}{4\sqrt 5}
  \left( \begin{array}{ccccc}
      -8 + 12 \sin^2 \theta  &
      \sqrt{\frac{3}{2}} \sin 2\theta  &
      \sqrt{\frac{3}{2}} \sin 2\theta  &
      \sqrt{\frac{3}{2}} \sin^2 \theta  &
      \sqrt{\frac{3}{2}} \sin^2 \theta \\  
      \sqrt{\frac{3}{2}} \sin 2\theta  &
      2 - 3 \sin^2 \theta  &
      -3 \sin^2 \theta  &
      -3 \sin 2\theta  &
      3 \sin^2 \theta \\  
      \sqrt{\frac{3}{2}} \sin 2\theta  &
      -3 \sin^2 \theta  &
      2 - 3 \sin^2 \theta  &
      3 \sin^2 \theta  &
      -3 \sin 2\theta \\  
      \sqrt{\frac{3}{2}} \sin^2 \theta  &
      -3 \sin 2\theta  &
      3 \sin^2 \theta  &
      2 - 3 \sin^2 \theta  &
      3 \sin 2\theta \\  
      \sqrt{\frac{3}{2}} \sin^2 \theta  &
      3 \sin^2 \theta  &
      -3 \sin 2\theta  &
      3 \sin 2\theta  &
      2 - 3 \sin^2 \theta \\  
 \end{array} \right),  \\
  \label{eq:l2hu2cpl}
  \MC C^{\theta}(2;H_u H_u;H_g)&=&
  \frac{1}{4}
  \left( \begin{array}{ccccc}
      0 &
      \sqrt{\frac{3}{2}} 
      \sin 2\theta  &
      \sqrt{\frac{3}{2}} \sin 2\theta  &
      - \sqrt{\frac{3}{2}} \sin^2 \theta   &
      - \sqrt{\frac{3}{2}} \sin^2 \theta  \\  
      \sqrt{\frac{3}{2}} \sin 2\theta  &
      2 - 3 \sin^2 \theta  &
      \sin^2 \theta  &
      \sin 2\theta  &
      \sin^2 \theta \\  
      \sqrt{\frac{3}{2}} \sin 2\theta  &
      \sin^2 \theta  &
      2 - 3 \sin^2 \theta  &
      \sin^2 \theta  &\sin 2\theta \\ 
      - \sqrt{\frac{3}{2}} \sin^2 \theta    &
      \sin 2\theta  &
      \sin^2 \theta  &
      -2 + 3 \sin^2 \theta  &
      \sin 2\theta \\  
      -\sqrt{\frac{3}{2}} \sin^2 \theta   &
      \sin^2 \theta  &
      \sin 2\theta  &
      \sin 2\theta  &
      -2 + 3 \sin^2 \theta \\ 
  \end{array} \right) .
\end{eqnarray}
In equation~(\ref{eq:secondorderapp}) the product
$\MC C^{\theta}(xx';T_{1u})^{\textrm{T}}\, \MC C^{\theta}(xx';T_{1u})$ 
enters. For the matrices given in~(\ref{eq:l1t1ucpl}), 
(\ref{eq:l1hgcpl}) and (\ref{eq:l1ggcpl}) these products can be expressed in
terms of the $l=2$ coupling matrices~(\ref{eq:l2t1ucpl}), 
(\ref{eq:l2hu1cpl}) and (\ref{eq:l2hu2cpl}):
\begin{eqnarray}
  \MC C^{\theta}(T_{1u}T_{1g};T_{1u})^{\textrm{T}}\, 
    \MC C^{\theta}(T_{1u}T_{1g};T_{1u})&=&
    \frac{1}{3}+\frac{1}{\sqrt 6}\,\MC C^{\theta}(T_{1u}T_{1u};H_g), \\
  \MC C^{\theta}(H_u H_g;T_{1u})^{\textrm{T}}\, 
    \MC C^{\theta}(H_u H_g;T_{1u}) &=&
    \frac{1}{5}+\frac{\sqrt 5}{10}\,\MC C^{\theta}(1;H_u H_u;H_g)
               -\frac{3}{10}\,\MC C^{\theta}(2;H_u H_u;H_g),  \\
  \MC C^{\theta}(H_u G_g;T_{1u})^{\textrm{T}}\, 
    \MC C^{\theta}(H_u G_g;T_{1u}) &=&
    \frac{1}{5}+\frac{\sqrt 5}{10}\,\MC C^{\theta}(1;H_u H_u;H_g)
               +\frac{3}{10}\,\MC C^{\theta}(2;H_u H_u;H_g).  
\end{eqnarray}
Using these relations, the  
total coupling matrix (neglecting the constant terms in the previous
relations) due to the applied $l=1$ and $l=2$ potential,
is given by 
\begin{eqnarray}
  \label{eq:LUMOcpl}
  H'(T_{1u},\theta)&=&\Big(V_{10}^2\,c_1+V_{20}\,c_2\Big)\;
     \MC C^{\theta}(T_{1u}T_{1u};H_g), \\
  \label{eq:HOMOcpl}
  H'(H_u,\theta)&=&V_{10}^2\,d_1\;
       \Big[\cos\delta_1\,\MC C^{\theta}(1;H_u H_u;H_g) +
        \sin\delta_1\,\MC C^{\theta}(2;H_u H_u;H_g) \Big] +  \\
    & &   V_{20}\,d_2\;
       \Big[\cos\delta_2\,\MC C^{\theta}(1;H_u H_u;H_g) +
        \sin\delta_2\,\MC C^{\theta}(2;H_u H_u;H_g) \Big] ,
\end{eqnarray}
where 
$c_1=\frac{1}{\sqrt 6}\,
  \frac{t(T_{1u}T_{1g};T_{1u})^2}{E_{T_{1u}}-E_{T_{1g}}}$ 
and  $c_2=t(T_{1u}T_{1u};H_g)$. Similarly we have
$d_1\cos\delta_1=\frac{\sqrt 5}{10}
  \Big[\frac{t(H_u H_g;T_{1u})^2}{E_{H_u}-E_{H_g}}+
  \frac{t(H_u G_g;T_{1u})^2}{E_{H_u}-E_{G_g}}\Big]$,
$d_1\sin\delta_1=\frac{3}{10}
  \Big[-\frac{t(H_u H_g;T_{1u})^2}{E_{H_u}-E_{H_g}}+
  \frac{t(H_u G_g;T_{1u})^2}{E_{H_u}-E_{G_g}}\Big]$ and
$d_2\cos\delta_2=t_1(H_u H_u;H_g)$, $d_2\sin\delta_2=t_2(H_u H_u;H_g)$.
Equation~(\ref{eq:LUMOcpl}) implies that the contributions of the 
$l=1$ and $l=2$ to the splitting of the LUMO add up trivially. 
Using the values in
\tab~\ref{tab:cc} and the energies of \fig~\ref{fig:c60Spectrum} yields the
values $\delta_1=0.064$ and $\delta_2=0.037$, which are almost equal when 
compared to
$\pi$. This can be understood by the following remarks: 
$\delta_1=0$ in the case of 
$t(H_u H_g;T_{1u})=t(H_u G_g;T_{1u})$ and $E_{H_g}=E_{G_g}$. Furthermore, it
can be shown that $\delta_2=\arctan(1/\sqrt{125})\approx 0.090$ assuming that the angular dependence of the
HOMO is given by $l=5$ spherical harmonics.
Taking an average value of $\delta=0.050$ yields the approximate relation
\begin{eqnarray}
  \label{eq:HOMOcpl2}
  H'(H_u,\theta)&\approx&\Big(V_{10}^2\,d_1+V_{20}\,d_2\Big)\;
       \Big[\cos\delta\,\MC C^{\theta}(1;H_u H_u;H_g) +
        \sin\delta\,\MC C^{\theta}(2;H_u H_u;H_g) \Big],  
\end{eqnarray}
which shows that, to a good approximation, the contributions of the 
$l=1$ and $l=2$ potential to the splitting of the HOMO add up trivially.

\section{Calculation of the matrix $\beta$} \label{sec:shtranslation}

In this section it is shown how to calculate the matrix $\beta$ appearing in
equation~(\ref{eq:vscr1}). The second term on the right side of this equation
describes the coefficients of the term  
$\sum_{\VC R_i\ne 0}V\tu i(\VC r-\VC R_i)$ which enters the screened
potential $V\tu{scr}$ and which describes the potential induced by all
neighboring sites. Using the definition~(\ref{eq:V}), we can rewrite this
expression as
\begin{equation}\label{eq:neighborpot}
  \sum_{\VC R_i\ne 0}V\tu i(\VC r-\VC R_i)=
    \sum_{lm} Q_{lm}\sum_{\VC R_i\ne 0}I^*_{lm}(\VC r-\VC R_i)=
    \sum_{lm} Q_{lm}\sum_{\VC R_i\ne 0}(-1)^m I_{l-m}(\VC r-\VC R_i).
\end{equation}
The function $I_{lm}(\VC r_1-\VC r_2)$ can  be decomposed  using the
translation formula (for $r_1<r_2$)
\begin{equation} \label{eq:shrel}
  I_{LM}(\VC r_1-\VC r_2)=
  \sum_{l_1,l_2=0 \atop l_2-l_1=L}^\infty (-1)^{l_2}
  \sqrt{\frac{(2l_2+1)!}{(2L\!+\!1)!(2l_1)!}}\,
  \sum_{m_1,m_2} C^{LM}_{l_1m_1\,l_2m_2}
    R_{l_1m_1}(\VC r_1)I_{l_2m_2}(\VC r_2),
\end{equation}
where $C^{LM}_{l_1m_1\,l_2m_2}$ denote the Clebsch-Gordan coefficients. 
This formula can be found
in different forms in the literature, see for example 
Ref.~\onlinecite{book:QToAngularMomentum,art:epton94}.
Substituting~(\ref{eq:shrel}) 
in the sum~(\ref{eq:neighborpot}) with
$\VC r_1=\VC r$ and $\VC r_2=\VC R_i$ yields
\begin{equation}\label{eq:neighborpot2}
  \sum_{\VC R_i\ne 0}V\tu i(\VC r-\VC R_i)=
    \sum_{l_1m_2\atop l_2m_2} \beta_{l_1m_1\,l_2m_2}
        Q_{l_2m_2} R_{l_1m_1}(\VC r),
\end{equation}
where the matrix $\beta$ is given by 
\begin{eqnarray} \label{eq:beta}
  \beta_{l_1m_1\,l_2m_2}&=&(-1)^{m_2+l_2}\,
  \sqrt{\frac{[2(l_1\!+\!l_2)]!}{(2l_1)!(2l_2)!}}\,
     C^{l_1\!+\!l_2\,m_1\!-\!m_2}_{l_1m_1\,l_2 -\!m_2}\,  
     \sum_{\VC R_i\ne 0}I_{l_1\!+\!l_2\, m_1\!-\!m_2}(\VC R_i)\\
     &=& (-1)^{m_2+l_2}\sqrt{\frac{(l_1\!+\!l_2\!+\!m_1\!-\!m_2)!
                     (l_1\!+\!l_2\!-\!m_1\!+\!m_2)!}
               {(l_1\!+\!m_1)!(l_1\!-\!m_1)!(l_2\!+\!m_2)!(l_2\!-\!m_2)!}}\,
    \sum_{\VC R_i\ne 0}I_{l_1\!+\!l_2\, m_1\!-\!m_2}(\VC R_i)
             \nonumber
\end{eqnarray}
In the last equality in~(\ref{eq:beta}) the explicit form of the
Clebsch-Gordan coefficients was used~\cite{book:QToAngularMomentum}. 
One verifies, that 
the matrix $\beta$ is complex conjugate under the exchange of all indices.
The remaining sums over the lattice sites $\VC R_i$ in~(\ref{eq:beta}) are
easily performed by computer. 
\end{widetext}

\end{document}